\documentclass[aip,amsmath,amssymb,preprint]{revtex4-1}%
\usepackage{graphicx}

\begin{document}
\title{Broadband negative refraction in stacked fishnet metamaterial}
\author{Zeyong Wei}
\affiliation{Department of Physics, Tongji University, Shanghai 200092, China}
\author{Yang Cao}
\affiliation{Department of Physics, Tongji University, Shanghai 200092, China}
\author{Jin Han}
\affiliation{Department of Physics, Tongji University, Shanghai 200092, China}
\author{Chao Wu}
\affiliation{Department of Physics, Tongji University, Shanghai 200092, China}
\author{Yuancheng Fan}
\affiliation{Department of Physics, Tongji University, Shanghai 200092, China}
\author{Hongqiang Li}
\email{hqlee@tongji.edu.cn}
\affiliation{Department of Physics, Tongji University, Shanghai 200092, China}

\begin{abstract}
We demonstrate a scheme to utilize the stacked fishnet metamaterial for
all-angle negative refraction and subwavelength imaging within a wide
frequency range starting from zero frequency. The theoretical predictions are verified
by the brute-force finite-difference-in-time-domain (FDTD) numerical
simulations. The phenomena come from the negative evanescent coupling between
the adjacent slab waveguides through the breathing air holes perforated on
metal layers.

\end{abstract}
\maketitle

Since J.B. Pendry proposed perfect lens\cite{1} using left-handed
materials\cite{2}, sustained attentions have been drawn to the negative-index
metamaterial (NIM) with simultaneously negative permittivity and permeability.
The NIM, comprising of subwavelength metallic resonant units, has been
designed and realized in both the microwave\cite{3} and optical regime\cite{4}%
. Negative refraction and subwavelength imaging with NIMs have great
application potentials in photonic devices\cite{5,6,7,8,9,10}. Among various
types of NIMs, one most promising candidate is the so-called fishnet NIM which
comprises of alternating metal/dielectric layers perforated with
two-dimensional array of holes\cite{11,12,13,14,15}. The simple structure also
provides a feasible solution for optical NIM\cite{16,17,18,19,20,21}.

In the previous studies on the fishnet
NIMs\cite{11,12,13,14,15,16,17,18,19,20,21}, the light waves are incident on
the top interface of the metal/dielectric multi-layers. Below the cut-off frequency of air holes the light waves can not penetrate into the structure and the negative index was retrieved within in a narrow frequency range above the cut-off. In this paper, a different incidence configuration is
employed by impinging the light waves on the sidewall interface of fishnet NIM
that is perpendicular to the metal/dielectric multi-layers. As the uniformly spaced holey metallic layers constitute a multiple of slab waveguides filled dielectric spacer layers, the incidence configuration of this kind enables us to fully exploit the optical properties of the fishnet NIM in the long wavelength limit. We show that the evanescent
coupling between the slab waveguides gives rise to all-angle negative
refraction and subwavelength imaging in a wide frequency range starting from zero frequency.

Figure 1 schematically illustrates the structure of our stacked fishnet
metamaterial and the incidence configuration. The metal/dielectric layers are
lying in $\hat{x}\hat{y}$ plane. The square arrays of air holes perforated on
metallic layers are aligned along z axis without lateral displacement in
$\hat{x}\hat{y}$ plane. The period of the hole array, the thickness of
metallic layer and dielectric layer are $p=6.0$mm, $t=0.035$mm and $h=1.575$mm
respectively. The line width of metallic strips along x direction $w=0.2$mm is
the same as that along $\hat{y}$ direction, and the size of square holes is
$a=p-g=5.8$mm. The dielectric constant of the dielectric layer is
$\varepsilon_{r}=2.2$. The EM incidence waves are propagating in the $\hat
{x}\hat{z}$ plane with an incident angle of $\theta$. \begin{figure}[ptb]
\includegraphics[width=8cm
]{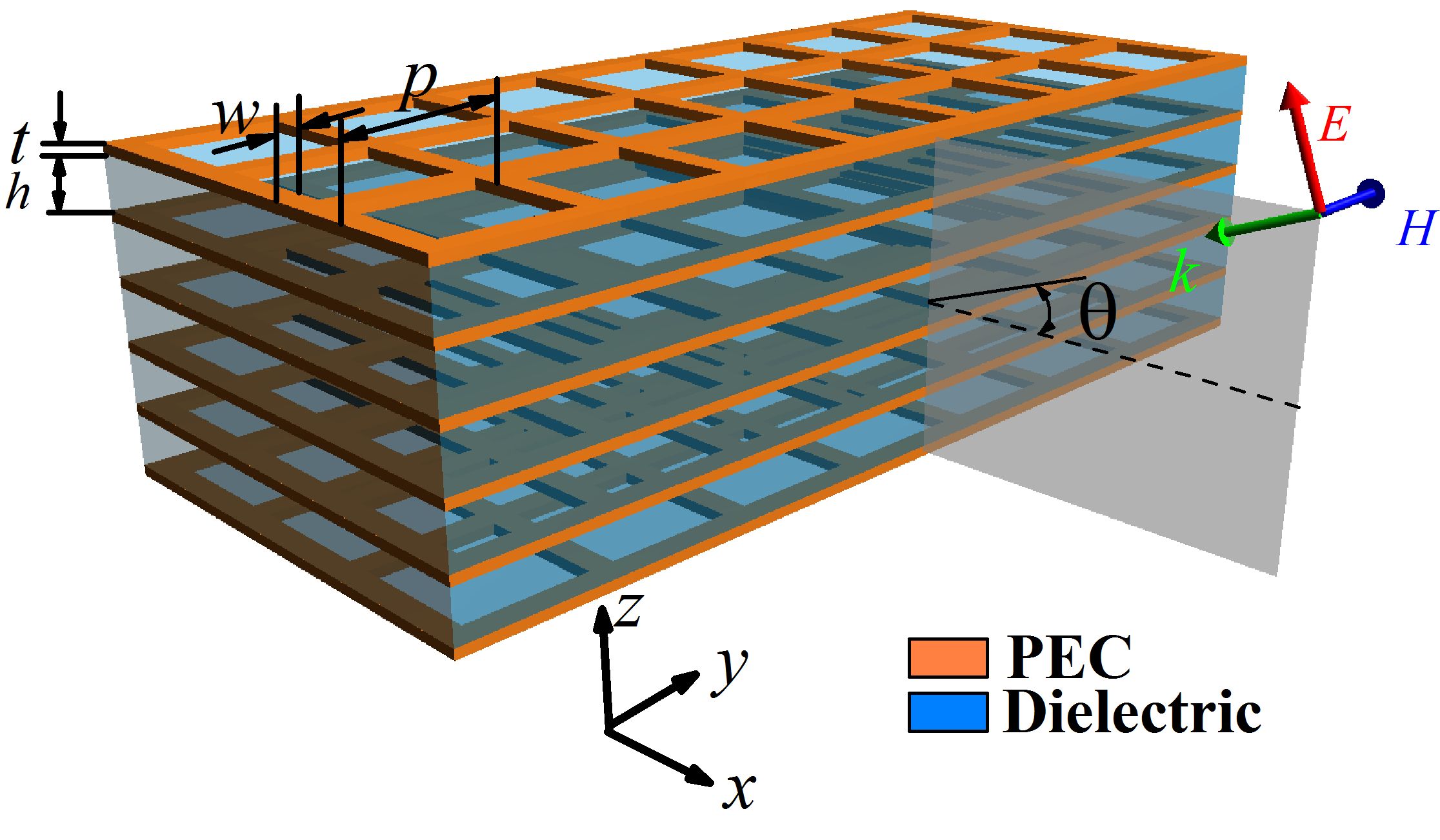}\caption{The schematic of stacked fishnet metamaterial. The red and
blue arrows indicate the directions of electric field $\vec{E}$ and magnetic
field $\vec{H}$ of incident waves. The plane in gray color denotes the incident plane.}%
\end{figure}
Under the assumption of perfectly electric conductor (PEC) for
metals, the electromagnetic fields in the region of holey metallic layers,
only existing in the air holes, shall be accurately expressed as the
superposition of local modes guided in square holes; while the waves inside a
dielectric spacer layer shall be expressed with the expansion of guided Bloch
waves in terms of periodicity in xy plane. By applying the boundary continuity
conditions at the interfaces of metal/dielectric layers (over the air holes)
and the periodic boundary condition along z direction for tangential
components of both the electric and magnetic fields, we can rigorously resolve
the dispersion relation of the three-dimensional stacked structure by the
modal expansion method$\cite{22,23,24,25}$. At wavelengths much longer than
both the thickness $h$ of dielectric layer and the size $a$ of square hole,
the calculations are quickly convergent with just one or a few more local
modes considered in hole waveguide. \begin{figure}[pb]
\includegraphics[width=8.3cm
]{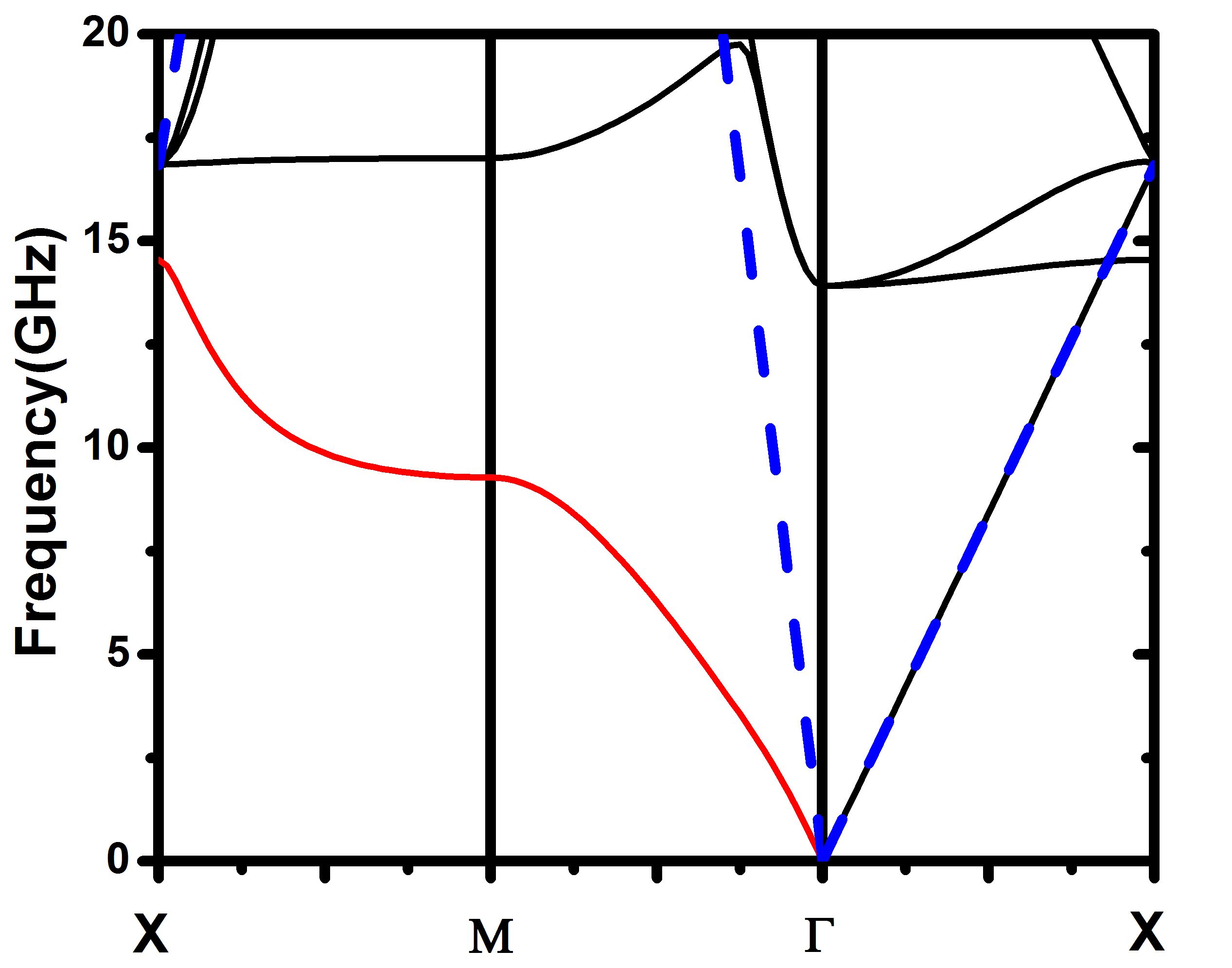}\caption{Dispersion diagram of the stacked fishnet
metamaterial along $\Gamma
$(0,0,0)$\rightarrow$\textrm{X}(0.5,0,0), $\Gamma$(0,0,0)$\rightarrow
$\textrm{M}(0.5,0,0.5) and M(0.5,0,0.5)$\rightarrow$X(0.5,0,0) directions.}%
\end{figure}

Figure 2 presents the calculated dispersion diagram along the $\Gamma
$(0,0,0)$\rightarrow$\textrm{X}(0.5,0,0), $\Gamma$(0,0,0)$\rightarrow
$\textrm{M}(0.5,0,0.5) and M(0.5,0,0.5)$\rightarrow$X(0.5,0,0) directions. The
blue dashed line refers to the light line in the dielectric. We notice that, the
lowest branch along the $\Gamma$X direction ($k_{z}=0)$ precisely coincides with
the light line. This is understandable with the help of modal
expansion method. Detailed calculations show that these $k_{z}=0$ states only
contain the 0$^{th}$ order Bloch component which is the transverse
electromagnetic (TEM) mode that is always orthogonal to the local modes of air
hole. In this situation, the free photons in the dielectric are the only
choice for the $k_{z}=0$ states as no evanescent couplings happen via the
breathing air holes. However the lowest branch ($k_{z}\neq0)$ along the
$\Gamma$\textrm{M} and \textrm{XM} directions (red solid line in Fig.2)
evidently deviates from the light line in dielectric. The $k_{z}\neq0$ states
on this branch originate from the evanescent coupling between the adjacent
slab waveguides via the TE$_{10}$ mode of holes (noting that the overlap
integral between a high order of guided Bloch mode and local mode of air hole
is not zero). Figure 3(a) presents the charts of equi-frequency surface (EFS)
analysis for the$k_{y}=0$ states to further reveal the characteristics of this
band. All curves in Fig. 3(a) are in a hyperbolic-like line shape, which
indicates that broadband all-angle negative refraction occurs in the $\hat{x}\hat{z}$
plane below the cutoff frequency of air holes. At lower frequency the
curves in Fig. 3(a) become much more flat, which means that the waves are also
strongly collimated inside the structure along the direction parallel to the
metal/dielectric layers. A numerical proof of negative refraction is shown in
Fig. 3(b). In our FDTD simulations, a monochromatic one-way Gaussian beam in
the $\hat{x}\hat{z}$ plane with a frequency at 11GHz is incident at an incident angle of 30$^{\text{o}}$. The fishnet model is stacked by 500 metal/dielectric layers along $z$ direction. Given the periodicity of hole arrays and the incidence configuration, 60 periods along x direction and one period along y direction are adopted for the metal/dielectric layers in the model. The negative refraction is clearly shown in Fig. 3(b) with the magnetic field distribution in the $\hat{x}\hat{z}$ plane. The black arrows denote the directions of energy flow in the free space and fishnet structure. A
refraction angle of -16.2$^{\text{o}}$, retrieved from the refracted direction
of the energy flow in the metamaterial or the negative Goos-Hanchen shift alternatively, is in good agreement with the estimate by EFS analysis. We also see from Fig. 3(b)
that the beam can easily propagate inside the structure without any reflection. \begin{figure}[ptb]
\includegraphics[width=8.3cm
]{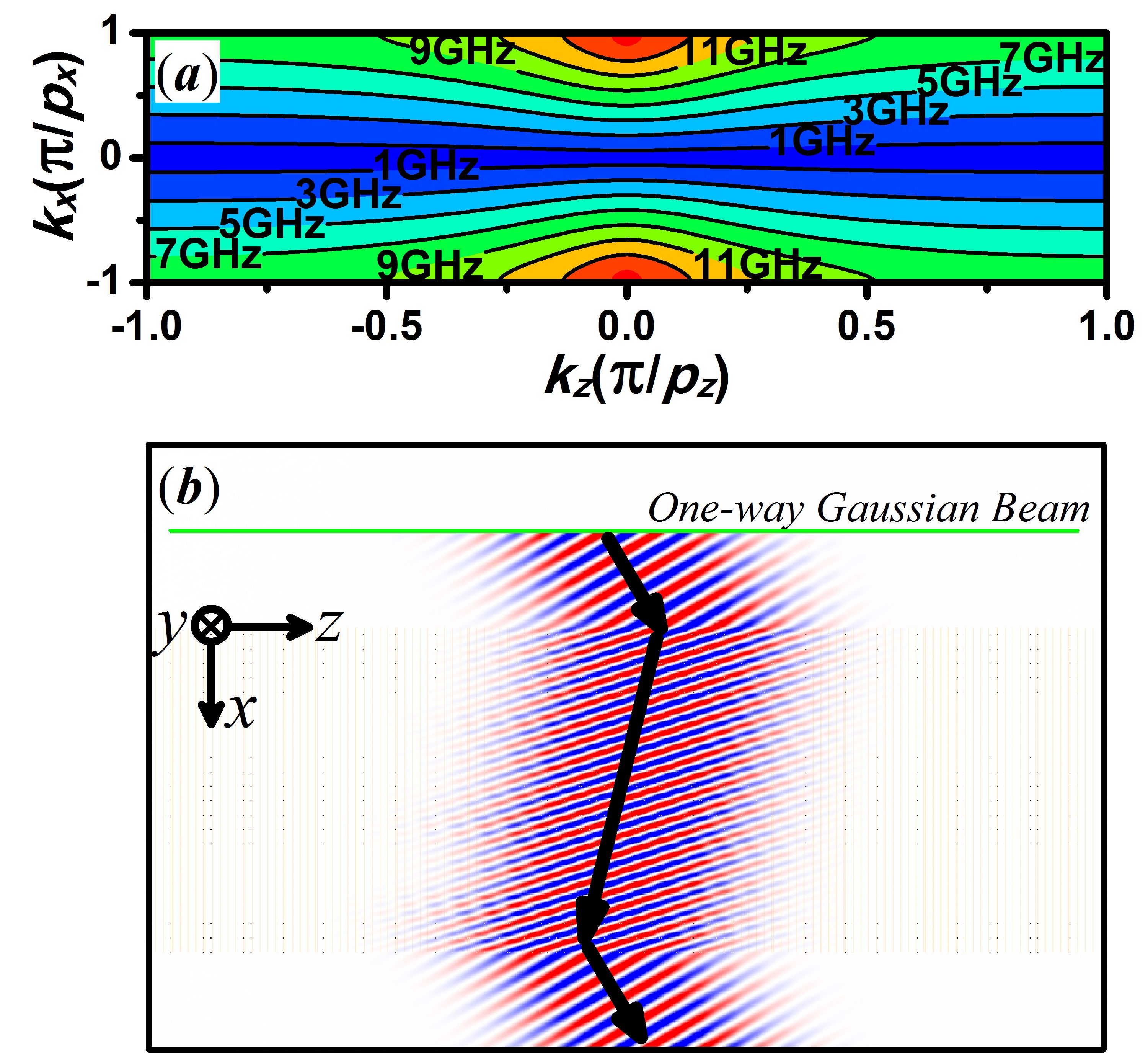}\caption{(a) The charts of EFS analysis for the lowest branch at ky=0.
(b) The distribution of magnetic field ($H_{y}$) in the $\hat{x}\hat{z}$ plane
calculated by the FDTD simulations. The green line denotes the position of the one-way
Gaussian beam about 70mm away from the top interface of our fishnet model.}%
\end{figure}

The optical properties of such a system can be described with the coupled wave
equation\cite{26,27,28} by considering the coupling between the $n^{th}$
waveguide channel and its nearest neighbors, the ($n-1)^{th}$, ($n+1)^{th}$
waveguide channels, as:
\begin{equation}
i\frac{da_{n}(x)}{dx}+\beta a_{n}(x)+C[a_{n+1}(x)+a_{n-1}(x)]=0\label{eq1}%
\end{equation}
Where $a_{n}(x)$ denotes the wave fields in the $n^{th}$ slab waveguide, C is
the coupling coefficient, and $\beta$ is the propagation constant of free
photons in the dielectric. Under the periodic boundary condition along z
direction, the dispersion of the system takes the form as
\begin{equation}
k_{x}=\beta+2C\cos(k_{z}p)\label{eq2}%
\end{equation}
where$\quad k_{x}$ and $k_{z}$ are the vector components along the $x$ and $z$
directions. At $k_{z}=0$, the coupling coefficient $C=0$ is zero (as
aforementioned no evanescent coupling occurs) and we have $k_{x}=\beta$ which
is rightly the light line in the dielectric. While at $k_{z}\neq0$, the coupling coefficient C is always negative in the limit of long wavelength (which can be deduced from the charts in Fig. 3(a)), giving rise to all-angle negative refraction. We note
that the silver/dielectric multi-layered structure also supports all-angle
negative refraction in a certain optical frequency regime under the same
incidence configuration of our study\cite{29}. The long range SPPs play an
important role for the negative refraction. We also note that, at long
wavelength limit, a holey metallic surface can be homogenized into a
single-negative medium with electric response in the form of Drude
model\cite{30}. The plasmon frequency is rightly the cut-off frequency of air
holes. Thus it is reasonable for us to consider the stacked fishnet
metamaterial as an artificial plasmonic waveguide array. The negative
coefficient C implies that for the eigenstates on the lowest branch, the
spatial field distributions are anti-symmetric with respect to the plane of
air holes. If our findings are applicable in the optical regime  where metals are dispersive and dissipative, the most field energy shall propagate outside the lossy metal film with anti-symmetric field distribution, and low loss is expected. The picture may be helpful to explain the low loss measured in a recent experiment about fishnet optical
NIMs\cite{16,19}. \begin{figure}[ptb]
\includegraphics[width=8.3cm
]{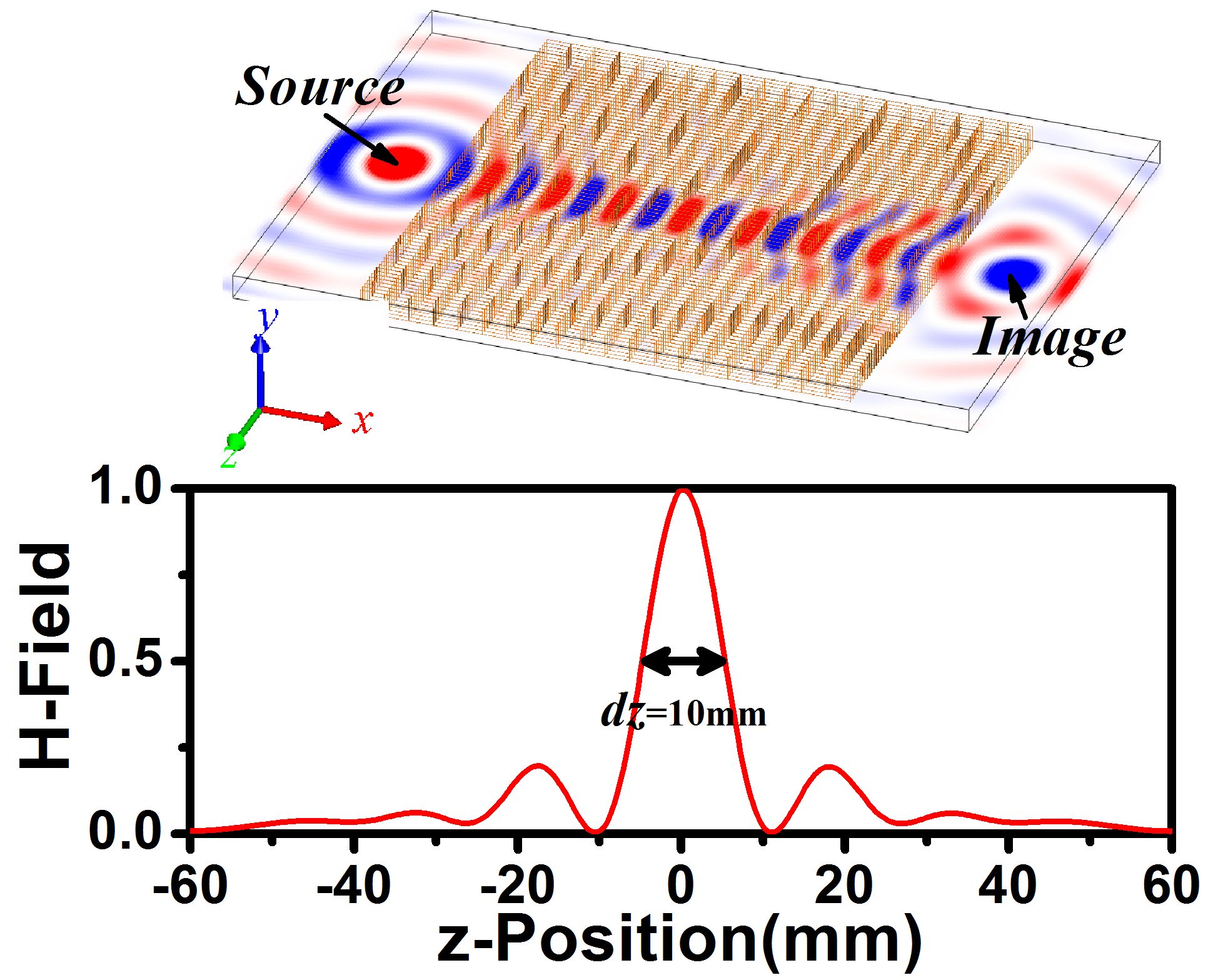}\caption{(a) The snap shot of magnetic field distribution in the
incident plane. The model, stacked by 80 metal/dielectric multi-layers along
z axis, has 20 periods along $\hat{x}$ direction and one period along $\hat{x}$ direction.
(b)Normalized magnetic field profile at the image plane as a function of z coordinate.}%
\end{figure}

One important application of all-angle negative refraction is flat lens. The
imaging performance of our stacked fishnet metamaterial is examined by the
brute-force FDTD numerical simulations. As shown in Fig. 4(a), a monochromatic
line source with a frequency at 11GHz is positioned 15mm away from the surface
at the left side of the fishnet structure. The snap shot shown in Fig. 4(a)
clearly indicates a high-quality image achieved at the image planeabout 15mm away from the outgoing interface at the other side. The image resolution can be
checked by the normalized magnetic field profile at image plane. As
illustrated in Fig. 4(b), along $\hat{z}$ axis, the full width at half maximum
(FWHM) of the field profile is 10mm about one-third of the wavelength. The
FWHM at a lower frequency still remains at about 10mm, leading to a better
resolution in subwavelength scale along the $\hat{z}$ direction. But a longer
structure along x direction is required due to stronger collimation effect at lower frequency. Further simulations indicate that the subwavelength imaging can be achieved in far-field (not shown).

In conclusion, the fishnet metamaterials can operate as plasmonic waveguide
arrays. Our findings about the broadband negative refraction and subwavelength imaging in the long wavelength limit have great potentials for photonic devices in microwave, THz and even in the optical regimes. This work was supported by NSFC (No.
10974144, 60674778), the National 863 Program of China (No. 2006AA03Z407),
NCET (07-0621), STCSM and SHEDF (No. 06SG24).

\end{document}